\documentclass[aps,prl,twocolumn,amssymb,amsmath,amsfonts,superscriptaddress,nofootinbib]{revtex4-1}

\usepackage{hyperref}
\usepackage{breakurl}

\usepackage{xcolor}
\hypersetup{
    colorlinks,
    linkcolor={red!50!black},
    citecolor={green!50!black},
    urlcolor={blue!80!black}
}

\def\beq{\begin{equation}}\def\eeq{\end{equation}}
\def\bea{\begin{eqnarray}}\def\eea{\end{eqnarray}}

\newfont{\cursive}{pzcmi at 9pt}

\newcommand{\LONGBIB}[2]{#2}

\newcommand{\AUTHAND}{and }
\newcommand{\ARXIVFULL}[1]{\LONGBIB{ [#1]}{}}
\newcommand{\CITELVC}{\LONGBIB{B.~P.~Abbott {\it et al.} [LIGO Scientific \AUTHAND Virgo Collaborations]}{LIGO Scientific \AUTHAND Virgo Collaborations}}
\newcommand{\CITEDOI}[1]{\LONGBIB{#1}{}}

\begin{document}

\title{Comments on:\\
``Echoes from the abyss:
Evidence for Planck-scale structure at black hole horizons"}

\author{Gregory Ashton}
\affiliation{Max-Planck-Institut f\"ur Gravitationsphysik, D-30167 Hannover, Germany}
\affiliation{Leibniz Universit{\"a}t Hannover, D-30167 Hannover, Germany}

\author{Ofek Birnholtz}
\email{ofek.birnholtz@aei.mpg.de.}
\affiliation{Max-Planck-Institut f\"ur Gravitationsphysik, D-30167 Hannover, Germany}
\affiliation{Leibniz Universit{\"a}t Hannover, D-30167 Hannover, Germany}

\author{Miriam Cabero}
\affiliation{Max-Planck-Institut f\"ur Gravitationsphysik, D-30167 Hannover, Germany}
\affiliation{Leibniz Universit{\"a}t Hannover, D-30167 Hannover, Germany}

\author{Collin Capano}
\affiliation{Max-Planck-Institut f\"ur Gravitationsphysik, D-30167 Hannover, Germany}
\affiliation{Leibniz Universit{\"a}t Hannover, D-30167 Hannover, Germany}

\author{Thomas Dent}
\affiliation{Max-Planck-Institut f\"ur Gravitationsphysik, D-30167 Hannover, Germany}
\affiliation{Leibniz Universit{\"a}t Hannover, D-30167 Hannover, Germany}

\author{Badri Krishnan}
\affiliation{Max-Planck-Institut f\"ur Gravitationsphysik, D-30167 Hannover, Germany}
\affiliation{Leibniz Universit{\"a}t Hannover, D-30167 Hannover, Germany}

\author{Grant David Meadors}
\affiliation{Max-Planck-Institut f\"ur Gravitationsphysik, D-30167 Hannover, Germany}
\affiliation{Leibniz Universit{\"a}t Hannover, D-30167 Hannover, Germany}
\affiliation{Max-Planck-Institut f\"ur Gravitationsphysik, D-14476 Potsdam-Golm, Germany}

\author{Alex B. Nielsen}
\affiliation{Max-Planck-Institut f\"ur Gravitationsphysik, D-30167 Hannover, Germany}
\affiliation{Leibniz Universit{\"a}t Hannover, D-30167 Hannover, Germany}

\author{Alex Nitz}
\affiliation{Max-Planck-Institut f\"ur Gravitationsphysik, D-30167 Hannover, Germany}
\affiliation{Leibniz Universit{\"a}t Hannover, D-30167 Hannover, Germany}

\author{Julian Westerweck}
\affiliation{Max-Planck-Institut f\"ur Gravitationsphysik, D-30167 Hannover, Germany}
\affiliation{Leibniz Universit{\"a}t Hannover, D-30167 Hannover, Germany}


\begin{abstract}
Recently, Abedi, Dykaar and Afshordi claimed evidence for a repeating damped echo signal
following the binary black hole merger gravitational-wave events
recorded in the first observational period of the Advanced LIGO interferometers.
We discuss the methods of data analysis and significance estimation leading to this claim,
and identify several important shortcomings.
We conclude that their analysis does not provide significant observational evidence
for the existence of Planck-scale structure at black hole horizons,
and suggest renewed analysis correcting for these shortcomings.
\end{abstract}

\maketitle

The detections by the Advanced LIGO detectors of gravitational wave signals from
binary black hole mergers
\cite{Abbott:2016blz,Abbott:2016nmj,TheLIGOScientific:2016pea} has opened up the
possibility of new tests of the nature of these objects
\cite{TheLIGOScientific:2016wfe,TheLIGOScientific:2016src,
TheLIGOScientific:2016pea}.
A recent work \cite{Abedi:2016hgu} has claimed
to find evidence of near-horizon Planck-scale structure using
data\cite{LOSC} from the three Advanced LIGO events
GW150914, LVT151012 and GW151226.
In the model of \cite{Abedi:2016hgu}
this near-horizon structure gives rise to
so-called echoes \cite{Cardoso:2016oxy, Cardoso:2016rao,Holdom:2016nek}.
Their inferred amplitude parameters suggest that
a lot of gravitational wave energy was emitted in the echoes:
a very rough calculation implies that the amount
of energy emitted in the echoes was approximately 0.1 solar masses (for GW150914)
and 0.2 solar masses (for LVT151012).
This should be compared to the total estimated energy emitted by the original signal
of 3 solar masses (for GW150914) and 1.5 solar masses (for LVT151012).
The data used is from the
LIGO Open Science Center (LOSC) \cite{LOSC}
which contains a total of 4096 seconds of
strain data from both Advanced LIGO detectors
around the three events.
Of these data the authors use only 32 seconds
centered around each event
for their analysis.
The authors claim to find such echoes
in data following the three events with
combined significance of 2.9$\sigma$
(p-value $3.7\times 10^{-3}$;
with the one-sided significance convention
used in \cite{Abbott:2016blz,Abbott:2016nmj,TheLIGOScientific:2016pea},
this value corresponds to 2.7$\sigma$).
If this claim were true,
it would force a major re-evaluation
of the standard picture of black holes
in vacuum Einstein gravity.

Besides the marginal claimed significance,
there are a number of aspects
of the analysis of \cite{Abedi:2016hgu}
that lead us to suspect that
the true significance of their detection
may be considerably weaker.
Here we will not examine the theoretical motivations
for the existence of such near-horizon Planck-scale structure,
nor the model templates the authors have chosen to search for.
Instead we will focus on the data analysis methods as reported
and the significance estimates assigned to them.
Regarding these we highlight some major data analysis drawbacks,
which cast doubt on this aspect of their result.

The first problem arises at how strong the relative signal should be for the
three events. The two binary black hole events GW150914 and GW151226 were
detected by the Advanced LIGO detectors with significance levels $>5.3\sigma$
and signal-to-noise ratios of $23.7$ and $13.0$
respectively\cite{TheLIGOScientific:2016pea}. The other event, LVT151012, had a
reported significance of only $1.7\sigma$ and a signal-to-noise ratio of $9.7$
combined between the two Advanced LIGO detectors.
However, in Table II of \cite{Abedi:2016hgu}
we see that the signal-to-noise ratio of
the claimed echo signal is actually largest for LVT151012.
The higher SNR cannot be due to
the projected number of echoes for LVT151012 over 32 seconds of data ($\sim\!180$)
being greater than the number of echoes for GW150914 over that duration ($\sim\!110$),
because
late echoes are strongly damped,
decreasing to a factor of 10 over $\sim\!22$ echoes.
Thus in order for the echoes of LVT15012 to have a higher 
SNR than the echoes of GW150914, 
their amplitude must be very high.
In fact to account for the reported SNRs,
the initial amplitude for the first echo of LVT151012
would have to be about $10\%$ higher than that of GW150914\footnote{
$\frac{\rho_{LVT151012}}{\rho_{GW150914}} = \frac{A_{LVT151012} \sqrt{\sum_{p=1}^{180} \gamma^{2p}}}{ A_{GW150914} \sqrt{\sum_{p=1}^{100} \gamma^{2p}}} = \frac{4.52}{4.13} \sim 1.1$,
where we have used the nomenclature of \cite{Abedi:2016hgu}, and $\gamma=0.9$.
},
while the original event's peak is about 2-3 times lower for LVT151012 in comparison to GW150914's.
This would require their parameter $A$ to be about 2-3 times larger for LVT151012 than for GW150914.
It would therefore be interesting to see plots and estimated parameters for LVT151012 (and GW151226),
similar to those presented in Table I of \cite{Abedi:2016hgu} for GW150914.

A second worrying aspect is the determination of the values for their echo
waveform model, Equation 9.
The model depends on six parameters:
a phase factor,
three time parameters $\Delta t_{\rm echo}$, $t_{\rm echo}$ and $t_{0}$, and two
amplitude parameters $A$ and $\gamma$.
The phase is modeled as a simple sign
flip at each reflection\footnote{ignoring the phase accumulated over the travel
between the light ring and the near-horizon Planck-scale boundary.},
$A$ is maximized over analytically,
and $\Delta t_{\rm echo}$ is determined by the parameters of
the final black hole given in \cite{TheLIGOScientific:2016pea} as given in
Equation 6.
The three parameters $\gamma$, $t_0$ and $t_{\rm echo}$
are determined by maximization,
with $\gamma$ and $t_0$ kept fixed between the different events.  
This maximization is done over a prior range, as
displayed in their Table I, and the values resulting from this maximization for
$\gamma$ and $t_{0}$ are found to lie very close to the boundary of this prior
range, $0.9$ and $-0.1$ respectively.
This suggests that there may be support
for values of these parameters that lie outside of this range
(no error ranges are given).
This would be particularly worrisome in the case of $\gamma$
since a value greater than unity means that
each successive echo would have an amplitude greater than the previous echo.
Such a result would seem unphysical, and if supported by the analysis method,
would cast considerable doubts on the method's validity.
Even if values $\gamma\geq 1$ are not supported,
the railing of the reported parameter values
against their prior range is 
a sign that these values may not be the best fits to the data;
if these values are in fact arbitrary,
reflecting the priors rather than the data,
they cannot be reliably considered as evidence for a detection claim.
It would be both helpful and prudent to show results of the analysis for wider prior ranges.

The third problem relates to how the background is estimated for their result,
as displayed in their Fig.~5.
For each time $t_{\rm echo}$ in a window 
covering offsets up to $\pm 5\%$ of $\Delta t_{\rm echo}$ after the merger, 
the matched filter SNR \cite{LOSCtutorial, Allen:2005fk}
is maximized over the remaining parameters 
$\Delta t_{\rm echo}$, $t_{0}$, $A$ and $\gamma$, either for GW150914 or for the
combined events.  In both cases 
the resulting peak of SNR is found to actually lie within $0.54\%$ of $\Delta 
t_{\rm echo}$.  They then estimate in each case how likely this peak would be in
random noise by finding how often such a high peak occurs in data away from the 
merger. However, since they originally allowed the time offset to range over 
$\pm 5\% \Delta t_{\rm echo}$, they should account for possible, comparable 
background peaks occurring over that full range 
not only the restricted range $(0-0.54\%)\Delta t_{\rm echo}$.
A na{\" i}ve accounting
for this post-hoc reduction in the extent of the parameter range 
would apply a trials factor of about 20 to the number of higher-SNR background 
samples, which would reduce the significance below $2\sigma$. 
A more sophisticated treatment of false positives over the reduced 
parameter range \cite{Connaughton:2016umz} indicates a trials factor of 
$\mathcal{O}(10)$, weakly dependent 
on the number of independent samples in the SNR time series.

It is unclear why this background estimation was performed using a
range of $t_{\rm echo}$ values that is only 9 to 38 echo periods away from the merger.
If there is indeed an echo signal in the data then this region
will not be entirely free of the signal being searched for.
At the beginning of the region the amplitude of the echoes
would only have dropped by a factor $0.9^9 \!\sim\! 0.4$.
One therefore expects a contaminated background estimation.
Each of the data sets released at \cite{LOSC} consist of 4096 seconds of data.
Both GW150914 and LVT151012 are located 2048 seconds into this data,
thus for large stretches of the data such contamination would be 
negligibly small. 
We expect that use of this relatively uncontaminated data would give a 
more self-consistent background estimate. 

A full analysis of the data is outside the scope of this comment.
Without a full analysis it is not possible to say whether the signals contain any 
true evidence of an echo signal,
but as discussed here there are sufficient
problems with the data analysis methodology of \cite{Abedi:2016hgu} to cast
grave doubt on their claimed significance of a $2.9\sigma$ effect.
It would be interesting to see the results of the analysis
with these problems addressed,
regarding both estimated parameters and estimated significance.
In conclusion, we find that the evidence as presented
in \cite{Abedi:2016hgu} is lacking in several key aspects,
such that their current methodology cannot
provide observational evidence for or against the existence of near-horizon
Planck-scale structure in black holes.

\begin{acknowledgments}
The authors thank Karl Wette, Francesco Salemi, Marco Drago, Andrew Lundgren
and Vitor Cardoso
for useful discussions,
and the authors of \cite{Abedi:2016hgu}
for helpful communications.
\end{acknowledgments}

\end{document}